\documentclass{emulateapj}

\usepackage{natbib}
\usepackage{graphicx, epsfig}
\usepackage{amssymb,ulem}

\usepackage[dvipsnames]{color}
\definecolor{webgreen}{rgb}{0,.5,0}
\definecolor{webbrown}{rgb}{.6,0,0}
\usepackage[pdfpagelabels]{hyperref}
\hypersetup{%
   colorlinks=true,hyperfootnotes=false,%
   breaklinks=true,%
   plainpages=false, bookmarksnumbered, bookmarksopen=true,
   bookmarksopenlevel=1,%
   urlcolor=webbrown, linkcolor=webbrown, citecolor=webgreen,
   }

\shorttitle{Galactic wind and $L_X$-SFR relation}
\shortauthors{Sarkar, Nath, Sharma \& Shchekinov}

\begin{document}


\newcommand{\3}{\ss}
\newcommand{\n}{\noindent}
\newcommand{\eps}{\varepsilon}
\def\be{\begin{equation}}
\def\ee{\end{equation}}
\def\bea{\begin{eqnarray}}
\def\eea{\end{eqnarray}}
\def\de{\partial}
\def\msun{M$_\odot$ }
\def\mpy{\msun yr$^{-1}$ }
\def\mmpy{{\rm M}_{\odot} {\rm yr}^{-1}}
\def\mpcc{m$_p$ cm$^{-3}$ }
\def\ergps{erg s$^{-1}$}
\def\div{\nabla\cdot}
\def\grad{\nabla}
\def\rot{\nabla\times}
\def\ltsima{$\; \buildrel < \over \sim \;$}
\def\simlt{\lower.5ex\hbox{\ltsima}}
\def\gtsima{$\; \buildrel > \over \sim \;$}
\def\simgt{\lower.5ex\hbox{\gtsima}}
\def\etal{{et al.\ }}


\title{Diffuse X-ray emission from star forming galaxies}

\author{Kartick C. Sarkar$^{1,2,\star}$, Biman B. Nath$^1$, Prateek Sharma$^2$, Yuri Shchekinov$^3$}
\affil{$^1$ Raman Research Institute, Sadashiva Nagar, Bangalore 560080, India}
\affil{$^2$ Joint astronomy Programme and Dept. of Physics, Indian Institute of Science, Bangalore, 560012, India}
\affil{$^3$  P. N. Lebedev Physical Institute, 53 Leninskiy Prospekt, 119991, Moscow, Russia}
\email{$^\star$kcsarkar@rri.res.in}

\begin{abstract}

We study the diffuse X-ray luminosity ($L_X$) of star forming galaxies using 2-D axisymmetric hydrodynamical simulations and analytical considerations of supernovae (SNe) driven galactic outflows. We find that the mass loading of the outflows, a crucial parameter for determining the X-ray luminosity, is constrained by the availability of gas in the central star forming region, and a competition between cooling and expansion. We show that the allowed range of the mass loading factor can explain the observed scaling of $L_X$ with star formation rate (SFR) as $L_X \propto$SFR$^2$ for SFR$\gtrsim 1$ \mpy, and a flatter relation at low SFRs. We also show that the emission from the hot circumgalactic medium (CGM) in the halo of massive galaxies can explain the large scatter in the $L_X-$SFR relation for low SFRs ($\lesssim$ few \mpy). Our results suggest that galaxies with small SFRs and large diffuse X-ray luminosities are excellent candidates for detection of the elusive CGM.
\end{abstract}


\section{Introduction}
\label{sec:intro}
Understanding the feedback mechanisms in galaxies is crucial in order to explain the  evolution of galaxies \citep{larson74, dekel86, msharma13} and enrichment of the intergalactic medium (IGM) \citep{tegmark93, nath97}. It has been observed \citep{strickland02, strickland07} and noticed in numerical simulations \citep[hereafter S15]{hopkins12, sarkar15a} that a significant fraction ($\sim 0.3-0.5$) of the input mechanical energy is stored in a hot ($T \gtrsim  10^6$ K), X-ray emitting gas. Therefore,  it is necessary to decipher the origin of diffuse X-ray emission from star forming galaxies to understand the feedback mechanisms.

In the case of stellar feedback processes producing a gaseous outflow, the hot gas can form in (i) the central region where star formation occurs, (ii) the free wind, (iii) the interaction zone between the wind and halo gas surrounding the galaxy, and (iv) the interaction region of wind and dense clouds \citep{suchkov94, suchkov96, strickland00, cooper08, cooper09, thompson15}. In addition, there is a non-negligible contribution from the hot halo gas surrounding the galaxies. For well-resolved galaxies, this basic scenario can be used to investigate the kinematic properties of the wind. For example, using X-ray observations, \cite{strickland07} found that the velocity of the outflow in the central region ($\sim 100$ pc) of M82 can be as large as $\sim 10^3$ km s$^{-1}$ and the mass outflow rate in the hot phase can be $\sim 1/3$ of the SFR in that galaxy. 

However, some aspects of the diffuse X-ray emission remain puzzling. Using 2D axisymmetric simulations for a galaxy with SFR $\sim 1$ \mpy , \cite{suchkov94} found that the shocked halo emission dominates over the emission from the central part. In contrast, using a full 3D simulation  of M82 (SFR $\sim 10 $ \mpy) \cite{cooper08} showed that most of the emission comes from the central region and free wind rather than halo.  \cite{cooper08, cooper09} also noticed that a part of the emission comes from the  interaction of clouds and the high velocity wind. However, a quantitative description of this emission is unavailable.

Another problem involves the scaling relation between the diffuse X-ray luminosity (not associated with point sources directly or indirectly) and the SFR. A thermally driven wind model 
  \citep[hereafter, CC85]{cc85}\footnote{Note that, the CC85 model with a smooth thermalised wind is only applicable for SFRs larger than a critical value ($\approx 0.1$ \mpy) \citep{sharma14}. Therefore, CC85 is a good approximation in the range of SFRs of our interest.} suggests that the hot gas density at the central region of galactic wind is $ \propto$ SFR, and therefore, the X-ray luminosity $\propto$ SFR$^2$. The temperature of the gas related to the wind or shocked halo is $\lesssim 2\times 10^7$ K which emits mostly in the soft band (0.5-2.0 keV).  A recent observational study of diffuse X-ray emission, however, suggests that the soft X-ray luminosity, $L_X, \propto$ SFR \citep[hereafter, M12]{mineo12}, though other scalings cannot be ruled out. \citealt{zhang14} and \citealt{bustard15} attempted to reconcile the observations with the expected scaling by adjusting parameters such as the mass loading factor (MLF; mass outflow rate/SFR $=\beta$) and the thermalisation efficiency ($\alpha$).
They suggested an inverse dependence of $\beta$ on SFR in order to explain the observed $L_X-$SFR relation. However, the physical origin for such an inverse relation remains  unexplained. 

Yet another problem is that galaxies with low SFR ($\le$ few M$_\odot$ yr$^{-1}$) show a flatter $L_X-$SFR relation with large scatter in the diffuse X-ray luminosity \citep[hereafter, W15]{wang15}, implying that other factors beyond stellar feedback contribute significantly to X-ray emission.

In this Letter, we constrain the mass loading factor based on the amount of interstellar medium (ISM) mass available and by the requirement that the cooling time be longer than the outflow expansion time.
Using this, we show that at large SFRs the X-ray luminosity ($L_X$) indeed scales as SFR$^2$, but at smaller SFRs the X-ray emission from the circumgalactic medium (CGM; which is insensitive to SFR) starts to dominate. This behaviour can lead to the observed $L_X \propto$ SFR or even flatter relation if one fits a single power law to observations.

%
%
%
%

\begin{figure}
\centering
\includegraphics[width=0.4\textheight]{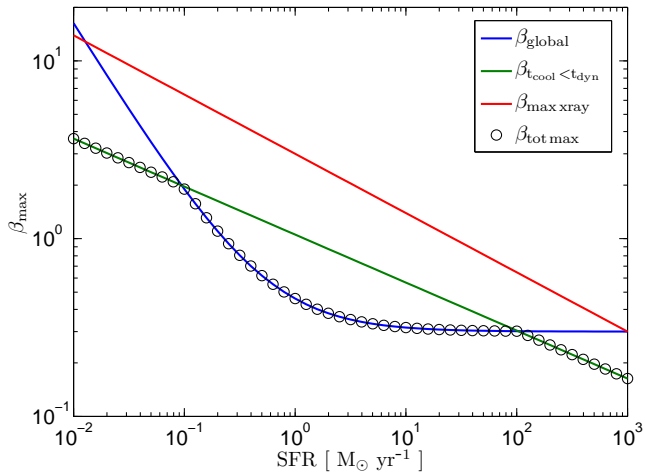}
\caption{Threshold values of MLF ($\beta$) as a function of SFR from various considerations: the availability of gas mass (blue); cooling time (green); and X-ray luminosity (0.5-2.0 keV) in the central region (red). The maximum allowed $\beta$ is shown by circles. Parameters used here are $R=200$ pc, $n_{\rm ism} = 10$ cm$^{-3}$, $\Delta t = 30$ Myr, $\alpha = 0.3$ and metallicity $=Z_\odot$. 
}
\label{fig:maxbeta}
\end{figure}
\section{Mass loading of outflows}
\label{sec:maxbeta}
Consider galaxies with outflows driven by thermal feedback from star formation, which we 
model as a thermal wind within a central region of size $R$ (following CC85).
The energy and mass injection in the central zone is parametrised by $\dot{M}$ and $\dot{E}$, which are respectively the mass deposition rate and the energy deposition rate, and are given by $\dot{M} = \beta$ SFR and $\dot{E} = 5\times 10^{15}\alpha$ SFR (assuming a Kroupa/Chabrier mass function, and an efficiency $\alpha \approx 0.3$ for energy deposition; here $\dot{E}$, $\dot{M}$ and SFR are in CGS units).

The X-ray luminosity of a galactic wind sensitively depends on the MLF $\beta$ (\citealt{zhang14}), which is governed by following considerations:
(a) Stellar evolution models suggest that stellar winds and supernova ejecta (without entrainment from the surrounding ISM) contribute to $\beta_0\approx 0.3$ \citep{leitherer99},
(b) the outflowing gas entrains mass from the surrounding ISM. However, the entrained mass (due to conduction and KH instabilities) cannot be larger than the total ISM mass $M_{\rm g} ( = 4\pi \mu  m_pn_{\rm ism} R^3/3)$ available within the central starburst region of radius $R$.
Therefore, an upper limit of MLF is given by
\be
\beta _{\rm global}= \beta_0+\frac{M_{\rm g}/\Delta t}{\rm SFR}=0.3+0.06\times \frac{n_{\rm ism}\: R_{\rm 100 pc}^3}{{\rm SFR}_{{\rm M}_\odot\: {\rm yr}^{-1}} \Delta t_{\rm Myr}}\,,
\label{eq:beta_global}
\ee
where, $n_{\rm ism}$ is the ambient ISM number density, $\Delta t$ is the age of the starburst .
(c) A further constraint arises from the cooling time of this central gas to be longer than the expansion time, otherwise most mass will condense radiatively and drop out of the outflow 
(see equation 10 of \citealt{thompson15};  for the curve shown in figure \ref{fig:maxbeta}, we use a wind opening angle of $60^\circ$). (d) A related constraint is that for the total X-ray luminosity of the central region ($\approx 4\pi n_c^2 \Lambda (T_c) R^3/3$; where, $n_c=0.3\dot{M}^{3/2}\dot{E}^{-1/2}R^{-2}/\mu m_p$ is the central ISM number density; $\mu=0.6$, is the mean molecular weight; $\Lambda$ is the X-ray emission function (\ergps cm$^{3}$); $T_c=1.4\times 10^7 \alpha/\beta$, is the central temperature (see CC85)) should be smaller than the energy deposition rate ($\dot{E}$), 
This gives an upper limit on MLF, namely,
\be 
\beta_{\rm max\,xray} = \left( \frac{13.5\,\alpha_{0.3}^2\,R_{100pc}}{{\rm SFR}_{{\rm M}_\odot\: {\rm yr}^{-1}}\Lambda_{-23}(T,Z)}\right)^{1/3}\,,
\label{eq:beta_xray}
\ee
where, $\Lambda_{-23}(T,Z)$ is the emission function at a particular X-ray energy band (in units of $10^{-23}$ \ergps cm$^3$), temperature ($T$) and metallicity ($Z$). For the calculation of $\beta_{\rm max\,xray}$ in figure \ref{fig:maxbeta}, we fix $\Lambda_{-23} (T_c, Z_\odot) = 1$. Note that argument (d) is not completely independent of argument (c).

In the case of high $\beta$, the outflowing gas has a large ram pressure ($\propto \dot{M}^{1/2}\dot{E}^{1/2} \propto \beta^{1/2}$) on the surrounding gas, 
and is likely to entrain more gas. It is therefore reasonable to assume that $\beta$ is likely to attain the maximum allowed value under the above considerations (b,c,d in previous paragraph).
Figure \ref{fig:maxbeta} shows various threshold values of $\beta$ as a function of SFR.
Open circles show the maximum values of $\beta$ allowed by these considerations.

\begin{figure*}
\centering
\includegraphics[width=0.8\textheight]{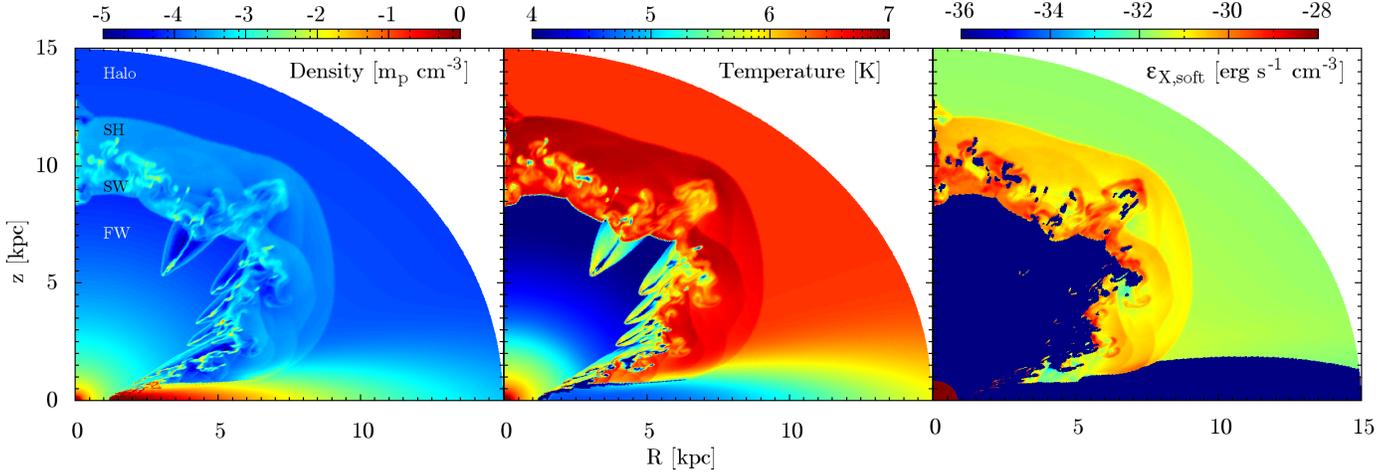}
\caption{Snapshots of density (left panel), temperature (middle panel) and soft X-ray (0.5-2.0 keV) emissivity (right panel) contours at $t = 20$ Myr for SFR = $5$ \mpy and central halo density $\rho_{h0} = 3 \times 10^{-4}$ \mpcc with total grid points $ = 512^2$. The labels in the left panel are as follows: FW- Free Wind, SW- Shocked Wind and SH- Shocked Halo. Note that we have used colourbar between $10^{-36}$ and $10^{-28}$ \ergps cm$^{-3}$ (right panel) but the core emissivity is $\sim 10^{-21}$ \ergps cm$^{-3}$.
}
\label{fig:snaps}
\end{figure*}
\section{Simulation details}
\label{sec:sim_details}
We perform 2-D axisymmetric hydrodynamic simulations using {\tt PLUTO} \citep{mignone07}. We simulate only one quadrant of  a MW type galaxy (total mass $M_{\rm vir} = 10^{12}$ \msun). The initial condition for  the galaxy is in dynamical equilibrium with a warm, rotating disc ($T \sim 4\times 10^4$ K; with Solar metallicity) and a hot gaseous halo ($T = 3\times 10^6$ K; with 0.1 Solar metallicity) surrounding the galaxy. We vary SFR for the same disc/halo properties. The disc gas is not allowed to cool if it is not shocked/perturbed; i.e., unless $ \sqrt{v_r^2+v_{\theta}^2}  \geq 20$ km s$^{-1}$. Other details of the model can be found in S15.

The SNe energy is deposited continuously in form of thermal energy in a spherical region of radius $R$ at the centre of the galaxy. In reality, most of the SNe occurs in a low density medium created by the previous SNe explosions and stellar winds. To mimic this, we create an artificially low density medium ($10^{-2}$ \mpcc) at $t = 0$ for $r \leq R$ (in local pressure equilibrium with the region outside) and then deposit the SNe energy and mass (with $Z_\odot$) inside it. This also prevents artificial cooling loses due to lack of sufficient numerical resolution. For estimating the X-ray emission function ($\Lambda_X (T,Z)$), we use {\tt MEKAL} model at 0.2 and 1.0 $Z_{\odot}$ and linearly interpolate for all other metallicities (from 0.1 to 1.0 $Z_{\odot}$). 
\section{Results}
\label{sec:sim-results}
Figure \ref{fig:snaps} shows snapshots of density, temperature and soft X-ray emissivity for SFR $ = 5$ \mpy and background halo density $\rho_{\rm h0} = 3\times 10^{-4}$ \mpcc at $t = 20$ Myr. It shows a typical structure containing free wind, termination shock, shocked wind, shocked halo and un-shocked halo as labelled in the left panel \citep{weaver77}. The soft X-ray (0.5-2.0 keV) emissivity (rightmost panel) shows the origin of X-ray emission in a typical galactic wind. It shows that the soft X-ray emissivity of the central region is very high and is followed by shocked wind, shocked halo and halo region.

We find that the luminosity of the central region becomes constant after $t \gtrsim 1$ Myr (which is essentially the time to set up a steady wind at the centre for a constant mass and energy injection rate, and is given by the sound crossing time $( \propto R/\sqrt{\dot{E}/\dot{M}}\, )$ for the hot wind). Though the contribution of the outer parts (consisting of the shocked wind, shocked halo and the CGM) increases
 with time because of the increased volume of the shocked halo gas 
 and continuous energy pumping from the wind, the X-ray luminosity from the central injection region and the CGM dominates.  Here we  present analytic scalings of these components. 
  
 Following CC85, the central luminosity (for $r \leq R$) can be estimated in the case of a uniform density ($\rho_c = \mu m_pn_c$) central region (of volume $4\pi R^3/3$) as
$
L_{X,C} 
               = 1.3\times 10^{40} \alpha^{-1}\beta^3{\rm SFR}
               ^2 R_{\rm 100 pc}^{-1}\Lambda_{-23}(T, Z) 
               \,{\mbox \ergps}
\label{eq:lx-sfr}
$. However this is an overestimate since the density in the central region is not quite uniform. Results from our simulations are well fit by,
\be
\frac{L_{X,C}}{\mbox \ergps} \approx 3\times 10^{39} \alpha^{-1}\beta^3{\rm SFR}
^2 R_{\rm 100 pc}^{-1}\Lambda_{-23}(T, Z) \, .
\label{eq:lx-sfr-new}
\ee

The next important contribution towards X-ray emission comes from the CGM which contains a significant fraction of the missing baryonic mass, as seen in X-ray \citep{anderson11, bogdan12, dai12} and absorption studies \citep{bordoloi14, borthakur15}. 

The CGM density profile can be approximated as $n_0 \, (1+r/r_c)^{-3/4}$, with a core radius $r_c$ ($\approx 3$ kpc) and central density $n_0$ (see Figure 1 of S15). While this is clearly an approximation, the density values are not that different from estimates in the literature (e.g., \citealt{sharma12,fang13,gatto13}). If the CGM gas is spread over a length scale $r/r_c=x\gg 1$, then the X-ray luminosity can be expressed in terms of $M_{CGM}(=10^{10} M_{CGM, 10}$ M$_{\odot}$), the total CGM gas mass (we express the dependence of $L_{X,CGM}$ on the extent of the CGM in terms of $M_{CGM}$), as
$
L_{X,CGM}\approx 5.4 \times 10^{40} \, n_{0,-3}^{4/3} \, r_{c,3}\,\Lambda_{-23} \, M_{CGM, 10}^{2/3} \, {\mbox \ergps} \,,
$
where $n_0=10^{-3} n_{0,-3}$ cm$^{-3}$ and $r_c=3 r_{c,3}$ kpc.
However, our simulation results show that the actual luminosity from CGM is somewhat less than this, because of the approximation $(x\gg1$) used in arriving at it, and is better represented by,
\be
\frac{L_{X,{\rm CGM}}}{\mbox \ergps}\approx  8.6 \times 10^{39} \, n_{0,-3}^{4/3} \, r_{c,3}\,\Lambda_{-23}(T,Z) \, M_{\rm CGM, 10}^{2/3} \,.
\label{eq:cgm-lx}
\ee

\begin{figure}
\centering
\includegraphics[width=0.4\textheight, angle=0]{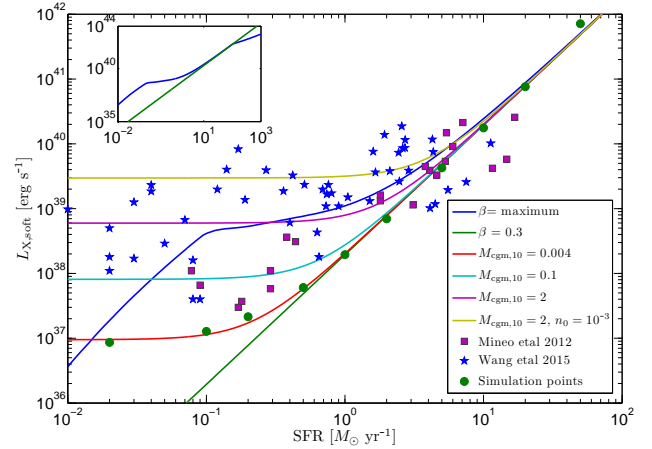}
\caption{Comparison of the soft X-ray luminosities and observed data (magenta squares for M12 and blue asterix for W15) for various models. The green and blue lines show $L_{X,C}$ for $\beta=0.3$ and $\beta_{\rm max}$ for $R=200$ pc. The inset shows that $L_X \propto$ SFR 
at larger SFRs ($\gtrsim 100$ \mpy) where cooling threshold becomes more important than mass loading (see Fig. 1).
The red, cyan and magenta lines show $L_{X}(=L_{X,C}(\beta=0.3)+L_{X,CGM})$ for $n_{0,-3}=0.3$, whereas, the golden line shows $L_{X}$ for $n_{0,-3}=1$. Notice that, high values of $L_X (\sim 10^{40} $ \ergps) for low SFR galaxies can be obtained for moderate values of $M_{\rm CGM}$ but using a higher value of $n_{0,-3}$ (also see eq. \ref{eq:cgm-lx}).
}
\label{fig:fit}
\end{figure}

Next we compare the X-ray luminosity from our simulations ( scaled according to Eqs \ref{eq:lx-sfr-new} and \ref{eq:cgm-lx} for different star formation and CGM properties) with the observed data.
Figure \ref{fig:fit} shows the $L_X$-SFR relation from our models. The green and blue lines show $L_{X,C}$ for the cases of  $\beta=0.3$ and the maximum $\beta$ ( circles in Figure \ref{fig:maxbeta}), respectively. We find that the data from M12, shown in red squares, are explained by $L_{X,C}$ for the range
of $0.3\le \beta \le \beta_{max}$, where $\beta_{max}$ is determined by the available ISM mass and the radiative cooling time, as discussed in section \ref{sec:maxbeta}, whereas, a higher $\beta$ for smaller SFRs due to the ISM mass loading makes the relation shallower at smaller SFRs. 
The data, which have hitherto been fit with a linear scaling between $L_X$ and SFR, actually belong to two different regimes: a quadratic scaling at large SFRs and a flattening at smaller SFRs.
In fact, the constraint of MLF from available ISM mass predicts $\beta \propto$ SFR$^{-2/3}$
from $0.1$ to a few \mpy (Figure \ref{fig:maxbeta}), which when put in eqn \ref{eq:lx-sfr-new} makes $L_X$ independent of SFR. 
Note that the ISM mass availability constraint, which we highlight for the first time, is the most stringent for SFRs of interest.

The different lines that flatten towards the lowest SFRs in Figure \ref{fig:fit} show the total luminosity (for $\beta = 0.3$) after adding the contribution from CGM with different masses and densities for $T = 3\times 10^6$ K and $Z=0.1Z_\odot$. We find that these curves can reasonably explain the data from W15 ( shown with the blue stars in Figure \ref{fig:fit}).
The X-ray luminosity in data flattens out at low SFRs because of the contribution from the CGM. W15 also find a dependence of $L_X$ on stellar mass; namely, $L_X/L_K \propto \left( {\rm SFR}/M_{\star}\right)^{0.3}$ (which is equivalent to $L_X/{\rm SFR} \propto [M_\star/{\rm SFR}]^{0.7}$ assuming $L_K \propto M_\star$), where $L_K$ and $M_{\star}$ are K-band luminosity and stellar mass of the galaxies, respectively. The stellar/halo mass dependence can naturally come from the CGM, which is more massive for larger galaxies (see Eq. \ref{eq:cgm-lx}). In fact, the relations above indicate a `fundamental plane' in $L_X,\,M_{\star}$ and SFR space, i.e. $L_X \propto M_{\star}^{0.7}$ SFR$^{0.3}$ (for SFR $\lesssim$ few \mpy), the existence of which can be tested with future observations.

Since the CGM mass is not yet reliably measured from observations, we can study the relation of $L_X$ with the expected scaling of CGM mass with stellar/halo mass. 
Recent observations suggest that about half of the missing baryons is in the form of cold clumps, and the rest could be warm-hot CGM gas \citep{werk14}. 
Since stellar mass comprises about a third, the mass of the warm-hot component of CGM gas can be comparable to $M_\ast$.

\begin{figure}
\centering
\includegraphics[width=0.4\textheight]{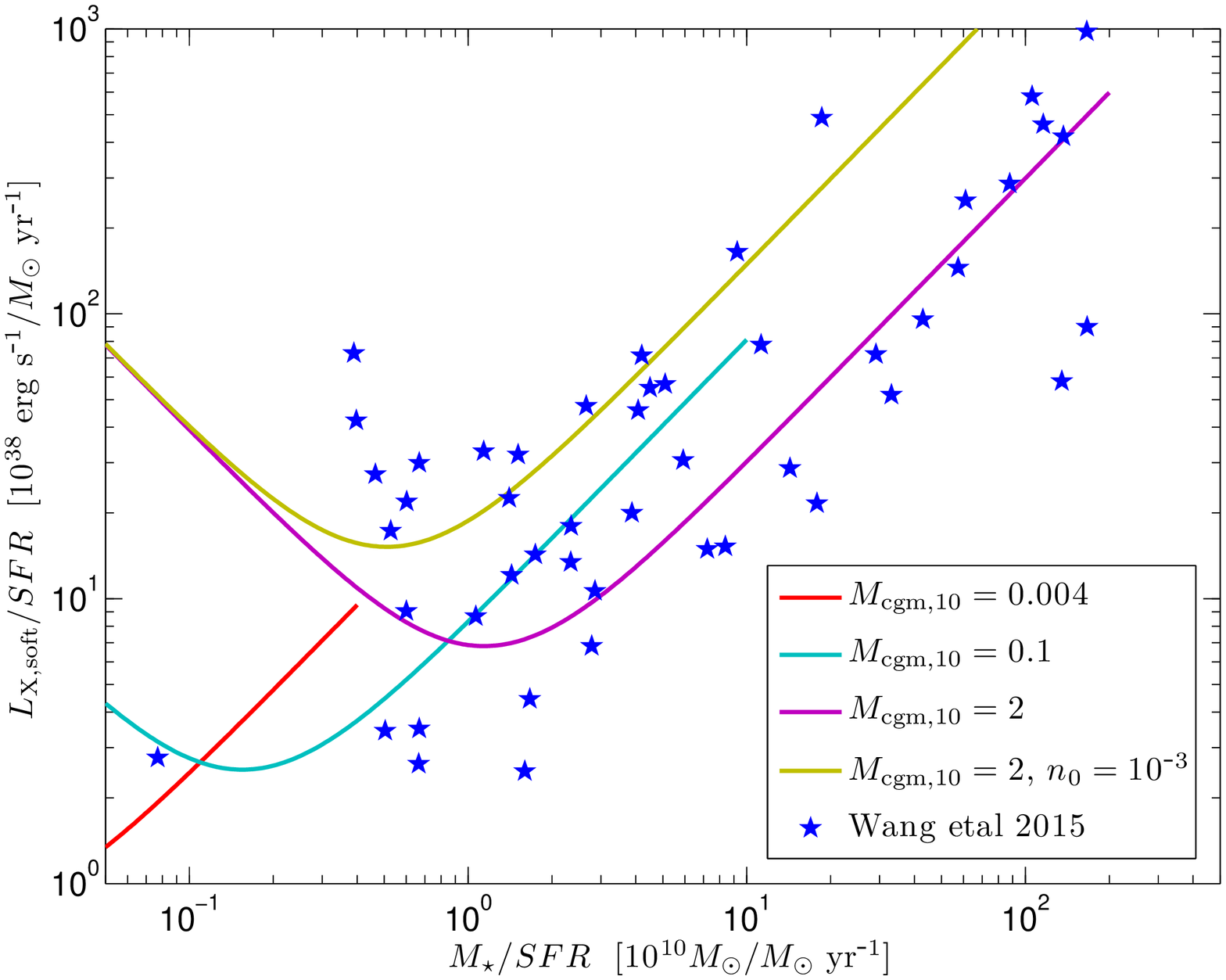}
\caption{Data from W15 along with curves for total diffuse X-ray luminosities for different values of $M_{CGM}$. These correspond to the same models as in Figure \ref{fig:fit} but normalised to $ M_{\rm CGM}=M_\star$.
}
\label{fig:fit2}
\end{figure}

Assuming that the CGM gas mass is equal to the total stellar mass,  in Figure \ref{fig:fit2} we show the relation between $L_X/$SFR and $M_\ast/$SFR, where $M_\ast$ is the total stellar mass. The data from W15 are shown along with the curves for different values of $M_{CGM} (=M_{\star})$. 
The highest SFR systems lie to left in this plot. The lines with different stellar/CGM masses look reasonably consistent with the data. 
The observed scaling of $L_X/$SFR $\propto (M_\ast/$SFR$)^{0.6}$ can be easily explained by the scaling of $L_{X,CGM}\propto M_{CGM}^{2/3}$ (eqn \ref{eq:cgm-lx}), for the CGM X-ray emission, which dominates in the low SFR (right portion of Figure \ref{fig:fit2}).
We also notice that the curves in Fig \ref{fig:fit2} show a negative slope for high SFR galaxies (on the left), which is consistent with the observed trend for high SFR galaxies in W15.

\section{Discussion}
\label{sec:discussion}
Our key result is that the diffuse X-ray emission from star-forming galaxies can be understood in terms of contributions from the central thermalised wind  (extending over $\sim$ 100 pc) and the extended CGM. For higher SFRs $L_X \propto$ SFR$^2$,
whereas, the CGM contribution dominates for SFR $\lesssim 1$ \mpy and accounts for the flattening of the $L_X-$SFR relation at low SFRs. Our model also predicts that the relation can be even flatter with a large scatter, depending on the halo properties. Since the CGM mass is expected to increase with the stellar/halo mass, at smaller SFRs a higher $L_X$ can result from the CGM contribution. 
In fact, the galaxies with low SFRs but high $L_X$ are likely to contain a large amount of CGM gas at temperatures of a few million degrees K, and are good candidates for spiral galaxies with a detectable X-ray emitting CGM (few such systems are reported by \citealt{anderson11,bogdan12}).

The X-ray luminosity from the CGM (eqn \ref{eq:cgm-lx}) depends on the CGM gas mass, density and temperature. We find that for the typical range in temperature (as found in, say, W15) of $2\hbox{--}8 \times 10^6$ K, the $L_{X,CGM}$ varies between $3 \times 10^{38}\hbox{--} 2.4 \times 10^{40}$ erg s$^{-1}$, for $M_{CGM}=10^{10}$ M$_\odot$. This spread arises from (a) the difference in emissivity with temperature and (b) the density profile of CGM gas at different temperatures. Figure \ref{fig:fit} shows that this spread in X-ray luminosity from the CGM gas can explain the data. We should however keep in mind that the spread in the data (Figure \ref{fig:fit2}) can partly arise from the spread in the relation between SFR and galaxy dynamical mass, which is likely related to $M_{\ast}$ \citep{karachentsev13}. We also note that the central SFR used in our models is an underestimation of a disc-wide SFR. This can also be responsible for the spread in the observed data.

It is generally believed that 
the CGM around low mass galaxies ($M_\star \lesssim$ few $\times 10^{9} M_\odot$) would have a low virial temperature (few $\times 10^5$K), which would make the CGM vulnerable to radiative cooling as  the cooling time would become less than the dynamical time of the galaxy \citep{singh15}. However, hot CGM around low mass galaxies can be formed from the hot and low density material ejected from disc supernovae which does not have sufficient energy to escape the galactic potential but have a long cooling time. This rejuvenated halo around low mass galaxies may give rise to the X-rays seen in low mass galaxies (which are also low SFR galaxies, in the presented data).
 The spread in $L_X-$SFR relation at the low SFR end can be partly due to the ill-understood, complex thermodynamic state of such low speed outflows. 

Though observations of the {\it total} X-ray emission (0.5-8.0 keV) \citep{mineo14} show a linear relation,  it is, however, supposed to be contaminated by high mass X-ray binaries (HMXB) \citep{grimm03} and should best be considered as an indicator to the SFR (since, number of HMXBs $\propto$ SFR)  rather than diffuse X-ray related to the galactic wind. 

We also note that the linear relation of X-ray luminosity from the shocked wind and halo as observed in highly inclined galaxies by \cite{strickland04b, tullmann06, li13} have to be studied separately as the soft X-ray emission from the central part of these galaxies is heavily absorbed by the galactic disc and does not represent the total emission. We will address these issues in detail in a future paper.

\section*{Acknowledgements}
We are indebted to Daniel Q. Wang for sharing data with us and for his useful comments. We also thank Nazma Islam for useful discussions. This work is partly supported by the DST-India grant no. Sr/S2/HEP-048/2012 and an India-Israel joint research grant (6-10/2014[IC]). YS acknowledges support from RFBR through 15-02-08293 and 15-52-45114, and partial support from the Grant of the President of RF for the Leading Scientific Schools NSh-4235.2014.2 .


\end{document}